\documentclass[12pt]{article}
\usepackage{color}
\usepackage{amsmath}
\numberwithin{equation}{subsection}

\numberwithin{equation}{section}

\begin{document}
\setcounter{page}{1} \pagestyle{plain} \vspace{1cm}
\begin{center}
\Large{\bf  Entropic Gravity and Cosmology in Kaniadakis Statistics}\\
\small \vspace{1cm}
{\bf N. Sadeghnezhad\footnote{Email: nsadegh@maragheh.ac.ir \  ,\   ORCID: 0000-0002-3629-5985}}\quad \\
\vspace{0.5cm} {\it Research Institute for Astronomy and
Astrophysics of Maragha (RIAAM) \\
University of Maragheh, P.O. Box 55136-553, Maragheh, Iran}

\end{center}
\vspace{1.5cm}
\begin{abstract}
             By using the Kaniadakis statistics, we discuss the modifications of Newtonian gravity and radial velocity profile in the light of Verlinde's formalism for gravitational entropy. After considering the implications of $\kappa$-statistics on the gravitational potential, it is shown that an accelerated universe may be obtained by considering the Friedmann first equation in this non- extensive statistics.\\
\end{abstract}
\vspace{1.5cm}
\newpage

\section{Introduction}
\quad The very deep and significant connection between gravitation, thermodynamics and quantum theory, have emerged since the notable discovery of black hole thermodynamics [1-6]. In this regard, Jacobson [7] derived the Einstein field equations starting from the Clausius relation at the horizon and using local Rindler observers. This pioneering work clearly indicates that the horizon entropy and the gravitational field equations are in one-to-one correspondence meaning that different entropies lead to distinct gravitational theory [8-9].
Indeed, in order to agree a black hole with the laws of thermodynamics, it is necessary to assign it an amount of entropy known as black hole entropy (or Bekenstein- Hawking entropy) [2,10-11]. In this way, the four laws of black hole mechanics [1-3,12] similar to those of thermodynamics, describe physical properties of black holes.\\
These laws and subsequent efforts showed that there is a deep connection between Einstein's gravitational field equations and the first law of thermodynamics [7,13]. Description of horizon area as a form of entropy and also surface gravity as the temperature, are evidences of this connection. This framework is also employable in the cosmological setups helping us to even justify the current and primary inflationary eras [14-23].   \\
Black hole entropy, according to the ''area theorem'' [24] and Bekenstein's conjecture [2,10,13], should be a monotonic function of area
\begin{equation}
S_{BH}=\frac{k_{B}A}{4l_{p}^{2}}=k_{B}\frac{Ac^{3}}{4G\hbar}\label{SBH}.
\end{equation}
Here, $A$ is the surface area of a black hole (area of the event horizon), $k_{B}$ is the Boltzmann constant and $l_{p}$ stands for the Planck length $\sqrt{\frac{G\hbar}{c^{3}}}$ while $G$, $\hbar$ and $c$ denote Newton's gravity constant, the Planck- Dirac constant and the speed of light, respectively.\\
For a stationary and spherically symmetric black hole, we can write
\begin{equation}
A=4\pi r_{h}^{2}=16\pi (\frac{GM}{c^{2}})^{2}\label{A},
\end{equation}
where $r_{h}=\frac{2GM}{c^{2}}$ denotes the horizon's radius and $M$ is the black hole's mass.\\
Bekenstein- Hawking (or Black- Hole) entropy in (\ref{SBH}) is also the maximal entropy that can be obtained by the Bekenstein bound [25]. Black hole physics and Bekenstein bound are strong supporting evidence for the holographic principle which states that microscopic degrees of freedom can be represented holographically either on the boundary of space- time or on horizons. Also, holographic bound limits how much entropy can be contained in matter and energy occupying a specified volume of space.\\
Considering a microscopic structure for gravity and assuming it as an emergent phenomenon, are the results that derived from the appeared close relationship between gravitation and thermodynamics. Soon after Padmanabhan introduced the statistical origin of gravity and connection between field equations of gravity and equipartion of energy [26], Verlinde addressed, in a holographic scenario, the non- fundamental beginning of gravity and suggested that gravity and space- time can indeed be explained as emergent phenomena [27]. In other words, the tendency of systems to raise their entropy is the origin of this emergence. In support of this approach, Verlinde regarded the gravity as a manifestation of entropy in the universe and derived Newton's gravitation law by this entropic force [27]. Staring from first principles, namely equipartition law of energy and using only space independent concepts, he shows that the Newton's law of gravitation appear naturally and also, a relativistic generalization of the arguments directly leads to the Einstein field equations of general relativity.\\
\\
In this framework, gravity is a macroscopic effect that arises from the changes of information regarding the positions of material bodies in the microscopic level, which is due to the tendency of systems in increasing their entropy [27]. In this regard, we can obviously assert the confirmed relevance of general relativity and thermodynamics in the black hole thermodynamics.  Verlinde's hypothesis has been so far a motivation for some works on cosmology, dark energy, cosmological inflation and cosmological acceleration [28-49].\\
According to Verlinde, the system tendency to enhance its entropy generates force, and thus the magnitude of this corresponding entropic force on a test particle which moves apart from a holographic screen is given by
\begin{equation}
F\Delta x=T \Delta S \label{FDX},
\end{equation}
where $\Delta x=\frac{\lambda_{c}}{8\pi}=\frac{\hbar}{8\pi mc}$, $T$ and $\Delta S=2\pi k_{B}$
are, respectively, displacement of test particle from the holographic screen, temperature and entropy change on the surface. Thus, a non-vanishing temperature is correlative with a non-zero force which, in turn, leads to a non-zero acceleration. According to Unruh [50], a temperature equal to
\begin{equation}
T=\frac{1}{2\pi k_{B}}\frac{\hbar a}{c} \label{T},
\end{equation}
will be experienced by an observer in a frame with acceleration $a$. $T$ in (\ref{T}) which is the necessary temperature to cause an acceleration $a$, can be considered as the temperature associated with the bits on the holographic screen.\\
Supposing a sphere with radius $R$ as a storage device for information, the number of used bits follows from
\begin{equation}
N=\frac{Ac^{3}}{G\hbar}\label{N},
\end{equation}
where $G$ and $A$ are Newton's constant and area of the holographic surface, respectively.\\
Assuming that the total energy of the system, $E$, is divided over the bits, temperature of the system can be determined by the equipartition rule
\begin{equation}
E=\frac{1}{2}Nk_{B}T \label{E}.
\end{equation}
The distributed energy on the screen $E$ is a reason for the presence of some mass $M=\frac{E}{c^{2}}$  surrounded by the screen. Thus, Verlinde recovered Newton's law of gravitation, practically from first principles [27]
\begin{equation}
F=G\frac{mM}{R^{2}} \label{F}.
\end{equation}
Boltzmann- Gibbs (BG) statistical mechanics and its related thermodynamics have certainly a fundamental position in the theoretical physics. Although statistical mechanics, as a formalism founded on the Boltzmann- Gibbs- Shannon entropy, can successfully describe the simple and short- range equilibrium systems, its validity and capability for complex physical, natural and artificial systems such as non-equilibrium systems and gravitational systems (which are long- ranged) is doubtful. So, it is useful to generalize the statistical mechanics to the non-extensive statistical mechanics which has the ability to overcome the limitations of BG entropy and describe some of the complex phenomena. Study of self- gravitating systems [51-56], solar neutrino deficit [57-60], peculiar velocities of galaxy clusters [61-62], flux of cosmic rays [63] and some cosmological setups [64-68] are samples for application of non- extensive statistical mechanics in astrophysics and cosmology.\\
The well- known Kaniadakis statistics [69-71], also known as $\kappa$-statistics among the other models, generalizes the standard BG statistics and can describe a large class of experimentally observed phenomena. This new entropy, which emerges from the relativistic generalization of the Boltzmann- Gibbs- Shannon entropy, generates power- law tailed statistical distributions which in classical limit reduce to the Maxwell- Boltzmann exponential distribution. The Kaniadakis statistics has been successful in the description of systems in the fields of gravity, cosmology, astrophysics, quantum physics among many others. This framework has succeeded when applied in various cases for example, analyzing the spectrum of the cosmic rays [72-73] and cosmic effects [74]. $\kappa$- entropy can be applied as a basis for a new model of holographic dark energy[75]. This scenario proves to lead to richer and more interesting cosmological behavior in comparison to usual holographic dark energy and is in agreement with observations [76-78]. Using the generalized Kaniadakis entropy and applying the gravity- thermodynamics conjecture, a new modified cosmology is obtained which leads to very interesting universe evolution [79] in agreement with observations [80].\\
In this paper, we apply the $\kappa$- statistics in the Verlinde's framework of gravitational entropy. Thereinafter, we consider modifications to the Newtonian gravity and implications of $\kappa$-statistics on the radial velocity profile and gravitational potential. We will see that it is possible to obtain an accelerated universe by considering the Friedmann first equation in the $\kappa$- framework.\\
This paper is outlined as follows. In section 2, we will provide a brief review of Kaniadakis formulation. In section 3, we employ the $\kappa$- statistics to derive the modified Newtonian second law of motion. In section 4, we will investigate some cosmological aspects of $\kappa$- statistics. The conclusions will be depicted in the last section.

\section{Kaniadakis statistics}
\quad Formally, the $\kappa$- framework is defined by one- parameter deformation of the exponential and logarithm functions as
\begin{equation}
\exp_{\kappa}(f)=(\sqrt{1+\kappa^{2}f^{2}}+\kappa f)^{\frac{1}{\kappa}},
\end{equation}
\begin{equation}
\ln_{\kappa}(f)=\frac{f^{\kappa}-f^{-\kappa}}{2\kappa},
\end{equation}
where deformation parameter $\kappa$ is an unknown parameter [69-71]. The constructed statistical mechanics recovers the ordinary Boltzmann- Gibbs statistical mechanics at the $\kappa \rightarrow 0$ limit.\\
The statistical distribution and entropy associated to this $\kappa$- framework can be written as

\begin{equation}
f_{\kappa}=\exp_{\kappa}(-\beta[E+\mu])\label{fk},
\end{equation}
\begin{equation}
S_{\kappa}=-k_{B}\sum_{i}^{W}\frac{p_{i}^{1+\kappa}-p_{i}^{1-\kappa}}{2\kappa}\label{Sk},
\end{equation}
where $\mu$ is the chemical potential and these relations reduce to the standard results in the limit $\kappa=0$. Specially, $\kappa$- entropy in (\ref{Sk}) leads to Gibbs entropy
\begin{equation}
S=-k_{B}\sum_{i=1}^{W}p_{i}\ln(p_{i}) \label{S},
\end{equation}
at the $\kappa\rightarrow 0 $ limit while $p_{i}$ is the probability for occupying the $i$th state and $W$ is the number of states in distribution. Non specified parameter $\beta$ in (\ref{fk}) contains all the information about the temperature of the system.\\
Using the kinetic foundation of $\kappa$- statistics [69-70, 81-82], the $\kappa$- equipartition theorem can be calculated through the relation
\begin{equation}
E_{\kappa}=\frac{1}{2}Nf(\kappa)k_{B}T \label{Ek},
\end{equation}
where $N$ is the total number of the degrees of freedom for the system and
\begin{equation}
f(\kappa)=\frac{(1+\frac{\kappa}{2})}{2\kappa(1+\frac{3\kappa}{2})}\frac{\Gamma(\frac{1}{2\kappa}-\frac{3}{4})\Gamma(\frac{1}{2\kappa}+\frac{1}{4})}{\Gamma(\frac{1}{2\kappa}+\frac{3}{4})\Gamma(\frac{1}{2\kappa}-\frac{1}{4})}
\quad 0\leq\kappa<\frac{2}{3}\label{f}.
\end{equation}
It is easy to see that the classical equipartition theorem for each microscopic degrees of freedom is recovered in the $\kappa=0$ limit ($ \lim_{\kappa\rightarrow0}S_{\kappa}=S_{BH} , E_{\kappa=0}=\frac{1}{2}Nk_{B}T $), while the expression (\ref{f}) diverges for $\kappa=\frac{2}{3}$.\\
In a microcanonical ensemble, which all the states have the same probability  $p_{i}=\frac{1}{W}$ , Gibbs entropy in (\ref{S}) leads to Boltzmann entropy
\begin{equation}
S=k_{B}\ln W,
\end{equation}
and using Bekenstein entropy (\ref{SBH}) we have
\begin{equation}
S_{BH}=k_{B}\frac{Ac^{3}}{4G\hbar}=k_{B}\ln W \label{SBH1}.
\end{equation}
In the so called ensemble, on the other hand, $\kappa$- entropy reduces to
\begin{equation}
S_{\kappa}=k_{B}\frac{W^{\kappa}-W^{-\kappa}}{2\kappa},
\end{equation}
where in combining with (\ref{SBH1})  and assuming $k_{B}=1$ , we have
\begin{equation}
S_{\kappa}=\frac{1}{\kappa}\sinh(\kappa S_{BH})=\frac{1}{\kappa}\sinh (\frac{A\kappa c^{3}}{4G\hbar}) \label{Sk1},
\end{equation}
found out previously in the [75].

\section{Dynamics in Kaniadakis gravity }
\quad Now, in the $\kappa$- statistics scenarios, the net applied force to particle $m$ by source $M$ is
\begin{equation}
F^{\kappa}=T \frac{dS_{\kappa}}{dA} \frac{\Delta A}{\Delta x} \label{Fk3},
\end{equation}
where by using (\ref{Sk1}) we have
\begin{equation}
\frac{dS_{\kappa}}{dA}=\frac{c^{3}}{4G\hbar}\cosh(\frac{A\kappa c^{3}}{4G\hbar}) \label{dSkdA}.
\end{equation}
Inserting Eq.s (\ref{T}), (\ref{N}) and (\ref{Ek}) in $E_{\kappa}=Mc^{2}$ we obtain
\begin{equation}
\frac{Ac^{3}}{4G\hbar}=\frac{\pi M c^{3}}{\hbar a f(\kappa)} \label{Ac3}.
\end{equation}
On the other hand, since $N$ is the number of bits we have $\Delta N=1$ from Eq. (\ref{N}) and thus
\begin{equation}
\Delta A=\frac{G \hbar}{c^{3}},
\end{equation}
which leads to
\begin{equation}
\frac{\Delta A}{\Delta x}=\frac{8 \pi G m}{c^{2}} \label{DADx}.
\end{equation}
Now, using Eq.s (\ref{T}), (\ref{dSkdA}), (\ref{Ac3}) and (\ref{DADx}), the applied force (\ref{Fk3}) can be written as
\begin{equation}
F^{\kappa}=ma^{\kappa}\cosh(\frac{\kappa}{f(\kappa)}\frac{\pi M c^{3}}{\hbar}\frac{1}{a^{\kappa}}) \label{Fk}.
\end{equation}
Considering that the total energy on the surface in (\ref{Ek}) is in fact the relativistic rest mass of the source mass $E=Mc^{2}$ and using Eq.s (\ref{T}) and (\ref{N}) we easily obtain
\begin{equation}
a^{\kappa}=G_{\kappa}\frac{M}{R^{2}}  \ : \   G_{\kappa}=\frac{G}{f(\kappa)} \label{ak},
\end{equation}
where $a^{\kappa}$ and $G_{\kappa}$ are the modified gravitational acceleration of test particle and effective gravitational constant in $\kappa$- framework, respectively.\\
Comparing $a^{\kappa}$ with Newtonian acceleration $a^{N}=\frac{GM}{R^{2}}$  , we see that they have the same sign because of the positivity of $f(\kappa)$ in the range of $\kappa$ . So, these two different frameworks expect a similar behavior for the acceleration.\\
Now, we can compare the modified gravitational force in $\kappa$- framework (\ref{Fk}), with Newtonian force and obtain
\begin{equation}
\frac{F^{\kappa}}{F^{N}}=\frac{1}{f(\kappa)} \cosh (\frac{\kappa \pi c^{3}}{\hbar G} R^{2})=\frac{1}{f(\kappa)} \cosh(\kappa \pi \frac{R^{2}}{l^{2}_{p}}) .
\end{equation}
where $l_{p}$ is the Planck length. This expression tends to unity at large distances, without any divergence, in the allowed range of $\kappa$. \\
In a circular motion, and using the acceleration $a^{\kappa}$ in (\ref{ak}), we easily reach at
\begin{equation}
v=\sqrt{\frac{G_{\kappa}M}{r}} \label{v},
\end{equation}
where $r$  and $v$ are respectively the radius and velocity of motion. $G_{\kappa}$ in (\ref{v}) must be positive, to have a real value for velocity, which that is the case for positive $f(\kappa)$ in the range $0<\kappa<\frac{2}{3}$.\\
The source mass $M$, also, has a gravitational field which applies the force $\frac{GM}{r^{2}}$  on the test particle $m$ . Equating (\ref{Fk}) with this force yields
\begin{equation}
\frac{GM}{r}=v^{2}\cosh(d\frac{r}{v^{2}})   :    d=\frac{\kappa}{f(\kappa)}\frac{\pi M c^{3}}{\hbar},
\end{equation}
which by expanding the $\cosh(d\frac{r}{v^{2}})$ and approximating it up to first order, we obtain the velocity profile as
\begin{equation}
v^{2}\cong\frac{GM}{r}-\frac{d^{2}}{2GM}r^{3} \label{v2}.
\end{equation}
The second term in rhs of (\ref{v2}) causes an decrease for the velocity of test particle in comparison to the Newtonian one.

\section{ Cosmological aspects in Kaniadakis framework  }
\quad Now, we want to calculate the Friedmann first equation in the acquired $\kappa$- gravity,  relying on an interesting approach to get the Friedmann equation (and the modified forms) from the Newtonian potential (and its modifications) [83]. Using the Eq.s (\ref{Fk}) and (\ref{ak}), one finds for the gravitational potential

\begin{eqnarray}
\phi(r)=\int\frac{F^{\kappa}}{m}dr=MG_{\kappa} \int \cosh(\frac{\kappa\pi c^{3}}{\hbar G} r^{2})\frac{dr}{r^{2}} \nonumber\\
=MG_{\kappa} \{\frac{\sqrt{\pi}}{2}\sqrt{\frac{\kappa\pi c^{3} }{\hbar G}}[erfi(\sqrt{\frac{\kappa\pi c^{3} }{\hbar G}}r)-erf(\sqrt{\frac{\kappa\pi c^{3} }{\hbar G}}r)]-\frac{\cosh(\frac{\kappa\pi c^{3}}{\hbar G}r^{2})}{r} \} \nonumber\\
=MG_{\kappa}[-\frac{1}{r}+\frac{1}{6}(\frac{\kappa \pi c^{3} }{\hbar G })^{2}r^{3}+O(r^{6})] ,
\end{eqnarray}
where $erf$ and $erfi$ are the error function and imaginary error function, respectively. Considering the high accuracy of Newton's theory in describing phenomena in weak gravitational fields, the deviation of the obtained result should be small compared to Newton's model. Therefore, the low orders of $\alpha$ and also the first order of $\kappa$ will be important, and despite the large $r$, expressions containing the second and higher orders of $\kappa$ can be omitted. Since this is a small correction, we expand the expression to the first order of $\kappa$, which corresponds to $r^{4}$. This potential can  be expanded up to $r^{3}$ as
\begin{equation}
\phi(r)\cong-\frac{MG_{\kappa}}{r}(1-\frac{\alpha^{2}}{6} r^{4}) \label{fi} ,
\end{equation}
where $\alpha=\frac{\kappa \pi c^{3}}{\hbar G}$ and $r$ is the location of apparent horizon. The integration constant ($C$) has been ignored since, whenever $\kappa=0$, the Newtonian potential ($\frac{MG}{r}$) is recovered only if we have $C=0$.    Following the used recipe in  [83] one can find some modified forms of the Friedemann equations by considering the Hubble law and various modified Newtonian potentials in writing the total energy of a test mass located at the edge of a FRW Universe. One of the well- known modified forms of the Newtonian potentials is
\begin{equation}
V(r)=(A+\frac{B}{r})V_{N}(r) \label{Vr},
\end{equation}
where $V_{N}=-\frac{GmM}{r}$ represents the Newtonian potential and $A$ and $B$ are unknown parameters which can be found by observational fitting, for example.
For a flat FRW universe and by fixing $A=1$ in Eq. (\ref{Vr}) one gets [83]
\begin{equation}
H^{2}=\frac{8\pi G}{3}\rho (1+BH) \label{H2},
\end{equation}
where $H$ and $\rho$ are the Hubble parameter and energy density of cosmic fluid, respectively.
Since we intend to regard the test particle on the edge of a flat FRW universe  ($r=\frac{1}{H}$), we use the (\ref{fi}) and (\ref{H2}) to find
\begin{equation}
H^{2}=\frac{8\pi G_{\kappa}}{3}\rho (1-\frac{\alpha^{2}}{6H^{4}}) \label{H21} .
\end{equation}
This modified Friedmann equation converts to standard one $H^{2}=\frac{8\pi G}{3}\rho$ in the limit $\kappa\rightarrow 0 (G_{\kappa}\rightarrow G)$. Also, in this limit, relation (\ref{H21}) can be written as below
\begin{equation}
H^{2}=\frac{8 \pi G_{\kappa}}{3} \rho (1-\frac{\rho_{c}^{2}}{\rho^{2}})  :   \rho_{c}^{2}=\frac{3 \alpha^{2}}{128 \pi^{2} G_{\kappa}^{2}},
\end{equation}
which can be analyzed in the form of a model similar to the cyclic universe.

\section{Summary }
\quad \\In this work, by considering the Kaniadakis statistics ($\kappa$-statistics) in the framework of Verlinde's hypothesis, we studied some modifications of Newtonian gravity. Applying of this non- extensive statistics to radial velocity profile shows an increase for the velocity of test particle in comparison with Newtonian one. Next, we explored the modification of gravitational potential in $\kappa$-statistics. Finally, we obtained the Friedmann first equation in a classical approach which results an accelerated universe and converts to standard one when the parameter of generalized entropy tends to zero. The consistency of the obtained results with observations is an interesting subject considered in future works.

\end{document}